\documentclass[fleqn,10pt]{article}
\usepackage{subcaption}
\usepackage{bm}% bold math
\usepackage{authblk}
\usepackage{graphicx}
\usepackage{amsmath,amsfonts,amssymb}
\usepackage[left=3.8cm, right=3.8cm, top=2cm]{geometry}
\pdfoutput=1
\usepackage{hyperref}
\usepackage{bm}% bold math
\usepackage{subcaption}
\usepackage[table]{xcolor}
\usepackage[symbol]{footmisc}

\title{On the use of multiple compartment epidemiological models to describe  the dynamics of influenza in Europe}
\date{\vspace{-5ex}}
\author[1,\footnote{inbarser@gmail.com}]{Inbar Seroussi}
\author[1]{Nir Levy}
\author[2]{Daniela Paolotti}
\author[3]{Nir Sochen}
\author[1]{Elad Yom-Tov}

\affil[1]{Microsoft, Herzeliya 46733, Israel}
\affil[2]{Computational Epidemiology Laboratory, Institute for Scientific Interchange, Turin, Italy}
\affil[3]{Department of Applied Mathematics, School of
	Mathematical Sciences,	University of Tel Aviv, Tel Aviv, Israel}
\begin{document}
\maketitle

\begin{abstract}
We develop a multiple compartment Susceptible-Infected-Recovered (SIR) model to analyze the
spread of several infectious diseases through different geographic areas. Additionally, we propose a data-quality sensitive optimization framework for fitting this model to observed data.

We fit the model to the temporal profile of the number of people infected by one of six influenza strains in Europe over $7$ influenza seasons. In addition to describing the temporal and spatial spread of influenza, the model provides an estimate of the inter-country and intra-country infection and recovery rates of each strain and in each season. We find that disease parameters remain relatively stable, with a correlation greater than $0.5$ over seasons and stains. Clustering of influenza strains by the inferred disease parameters is consistent with genome sub-types. Surprisingly, our analysis suggests that inter-country human mobility plays a negligible role in the spread of influenza in Europe. Finally, we show that the model allows the estimation of disease load in countries with poor or none existent data from the disease load in adjacent countries.

Our findings reveal information on the spreading mechanism of influenza and on disease parameters. These can be used to assist in disease surveillance and in control of influenza as well as of other infectious pathogens in a heterogenic environment. 
\end{abstract}

%\section*{Author summary}
%
%The seasonal spread of influenza is very complex due to the yearly invasion, extinction and subsequent invasion of multiple viral strains, as well as different spreading mechanisms among diverse populations groups. Previews work studied the global spread of a single strain of influenza or of influenza-like illness in multiple populations using epidemiological and proxy data. Here we develop a model which allows a study of the spatial and temporal spread of multiple disease strains and apply it to the speard of influenza in Europe. The model is fit to data via a novel optimization framework which allows consideration of all available data while automatically accounting for noisy and missing data. The product of the proposed model is a prediction of the spatial and temporal spread of influenza as well as the disease parameters. We show that influenza subtypes cluster according to their inferred parameters in a way that is consistent with genomic sub-types. Surprisingly, our analysis suggests that inter-country human mobility plays a negligible role in the spread of influenza in Europe. Our framework is general and can be applied easily to other diseases, models and data.  

\section{Introduction}
Understanding the spatio-temporal spread of infectious diseases is important for designing prevention mechanisms and control intervention strategies. The modeling of the dynamics of disease spread began with compartmental models, such as the Susceptible-Infected-Recovered (SIR) model \cite{kermack1932contributions}. This classic model, which describes a single pathogen in a heterogeneous environment, was later extended to cases of collections of interconnected geographic areas, spatial networks and multiple strains to model common infections such as influenza, measles, and foot-and-mouth disease \cite{charu2017human,levy2018modeling,chattopadhyay2018conjunction,kryazhimskiy2007state}. 

Empirical data, including both epidemiological and proxy data such as postings on social media, can be fit to these models to gain better understanding of an epidemic. These  models are also useful in simulating disease progression using known demographic variables and disease parameters \cite{colizza2006role,balcan2009multiscale}. For these two purposes multiple compartments models can account for heterogeneity in the population and disease \cite{riley2007large,hagenaars2004spatial}. 

Here we aim to model the seasonal spread of influenza in Europe. Seasonal spread of influenza is highly complex due to 
the yearly invasion, extinction and subsequent re-invasion of viral strains as well as the spatial spreading mechanism among different populations. Previous work \cite{colizza2006role,charu2017human} studied the global spatial and temporal spread of influenza of a single pathogen as well as that of influenza-like illness (ILI), as caused by human mobility patterns. These investigations showed the importance of air transportation and local traffic in the spreading mechanism. Similarly, \cite{brownstein2017combining} used modeling in conjunction with transportation data to generate surveillance systems and forecast ILI rates.   

However, a unified characterization of the underlying mechanisms of spread of multiple strains among multiple populations is missing. One of the reasons for this is the dearth of high resolution (spatial, temporal and biological) epidemiological data which has only recently become available. 

Here we develop a multi-compartment model to estimate the temporal and spatial spread of disease and apply it to modeling of influenza in Europe. The proposed model also provides and estimate of the infection and recovery rates. Fitting observed disease load data to such a model is a challenge, due to the complexity of the model and the availability of imperfect data. Thus, we develop a new optimization technique which allow us to take into account all available data while giving lower weight to noisy data during the optimization process. This framework allows us to add constraints on factors such as the distribution of infection and recovery rates. The model we consider applies to general networks and does not require simplifying assumption on the connectivity between countries, in contrast to models such as gravity models\cite{sattenspiel2009geographic}. 
After fitting the model to observed infection rates, we evaluate the estimated model parameters against known demographic data and virological information, gaining insights on the form of infection and recovery rates in different types of influenza. 

\section{The multi-compartment SIR model}

The model we develop is based on the classic SIR model, due to Kermack and McKendrick\cite{kermack1932contributions}. 
The model quantifies the number of people in each of three groups in a population: susceptible ($S$), infected ($I$), and recovered (or immune) ($R$). The evolution within these groups is described by a system of ordinary differential equations:
\begin{equation}
	\frac{dS}{dt} = - \beta S I , \quad
    \frac{dI}{dt} = \beta S I - \gamma I, \quad
    \frac{dR}{dt} = \gamma I
    \label{scalar_sir}
\end{equation}
where $\beta > 0$ is the infection rate and $\gamma > 0$ is the recovery rate.

Multiple viruses and groups of sub-populations can be accounted for \cite{levy2018modeling, seroussi2019multi} by transforming Eq. (\ref{scalar_sir}) to a multidimensional representation in which $S$, $I$, $R$, $\bm{\beta}$, and $\gamma$ are tensors (Eq. (\ref{matrix_sir})). 
These tensors act as a time-dependent state representation of the entire system, reflecting its complex multidimensional evolution, where an infectious sub-population interacts with other sub-populations under the influence of multiple viruses. Each two-dimensional projection of the tensors represents different sub-groups e.g. viruses, spatial regions. We refer to this model as a multi-compartment SIR (mcSIR).

Thus, the multi-compartment model is represented by a multi-dimensional set of equations. The dynamics of the mcSIR model is represented by the following system of ordinary differential equations:
\begin{equation}
	\begin{gathered}
    \begin{aligned}
		\frac{d \mathbf{S}}{dt} &= - \mathbf{I \bm\beta^\intercal S} \\
  	  	\frac{d \mathbf{I}}{dt} &= \mathbf{S \bm\beta^\intercal I - \bm\gamma^\intercal I} \\
  	  	\frac{d \mathbf{R}}{dt} &= \mathbf{\bm\gamma^\intercal I} \\
 	  	\label{matrix_sir}
	\end{aligned}
	\end{gathered}
\end{equation}
where boldface represents a tensor. This representation of sub-populations and viruses can also be generalized to other epidemic models, such as Susceptible-Infected-Recovered-Susceptible (SIRS) and Susceptible-Infected-Susceptible (SIS) models \cite{keeling2011modeling}. 

In the analysis below we consider a system of $N$ sub-populations and $V$ virus strains. For simplicity, we consider a diagonal model in the dimension of the strains, i.e., allowing no mutations or cross infections between virus strains. The off-diagonal elements are only with respect to infection rates between different population groups (e.g., countries). 

We note that in some cases multiple solutions of the mcSIR model can be fit to the same data. Specifically, an mcSIR model with zero off-diagonal $\bm{\beta}$ and $\gamma$, which are identical to the single-compartment SIR model can describe the same infected time series when the recovery rates of the two models are equal and when the sum of each row $\bm{\beta}$ is equal to the equivalent $\bm{\beta}$ in the diagonal model. Under these conditions (and assuming the same initial conditions) the two solutions are indistinguishable. In such a case, given only the time series of the infected population, the non-diagonal mcSIR solution represents high mixing which cannot be distinguished from the diagonal mcSIR solution, where no mixing occurs. 

\section{The Lagrangian Optimization Framework}

We propose a variational technique to match the observed number of infected people to the mcSIR model. The core of this technique is a scalar energy function, i.e, the Lagrangian, $L(\dot{\bm{S}},\bm{S},\dot{\bm{I}},\bm{I},\dot{\bm{R}},\bm{R})$. The stationary solution of the Lagrangian is defined by the Euler-Lagrange equations\cite{arnol2013mathematical}. The resulting Euler-Lagrange equations are a variation of our compartment model: The SIR model (Eq. (\ref{scalar_sir}) in the case of one compartment, and the mcSIR model (Eq. (\ref{matrix_sir})) in the case of multiple compartments. 

We add to this functional a constraint on the distance between the number of infected people as predicted by the model and the observed data. The Lagrange multiplier coefficients are learned during the optimization process. For example, in the single compartment model, the final estimated Lagrange multiplier is a weight showing to what extent the model fit provides a good representation of the data. 

To simplify the problem, we reduce the variational problem to one which accounts for only the susceptible and infected people. The number of recovered people can be computed from the number of infected people over time. In principle, the number of recovered people can also be incorporated in the Lagrangian. However, since the observed data usually only measures the number of infected people, using the reduced Lagrangian for $S$, and $I$ is sufficient. 

The reduced Lagrangian, $L_{\mathrm{SIR}}(\dot{S},S,\dot{I},I)$ of the SIR model in this case is then: 
\begin{multline}
    L_{\mathrm{SIR}}=L_{\mathrm{SIR}0}+\int^t_0\left[\frac{\dot{S}I-S\dot{I}}{2}+\left(\beta IS+u(t)\right)\left(I+S-S_{0}-I_{0}-\frac{\gamma}{\beta}\mathrm{log}(\frac{S}{S_{0}})\right)\right]\\+\frac{\lambda}{2}\int^t_0\left( I_{\mathrm{data}}-I\right)^2 dt+\lambda_2\mathrm{log}P(\beta,\gamma),
     \label{eq:LSIR}
\end{multline}
where $L_{\mathrm{SIR}0}$ is an initial condition, $\lambda$ is the Lagrange multiplier constraining the similarity to the data. The model parameters are the infection rate $\beta > 0$ and the recovery rate $\gamma > 0$. The coefficient $\lambda_2$ is an optional additional constraint on the distribution $P(\beta,\gamma)$ of the infection rate $\beta$ and recovery rate $\gamma$. This constraint is added if one is given information on disease parameters, for example, the range of values and their average. The Euler-Lagrange equations based on the functional in Eq. (\ref{eq:LSIR}), for the traditional SIR model, Eq. (\ref{scalar_sir}) are as follows: 
\begin{equation}
\begin{gathered}
\begin{aligned}
\dot{S}&=-\left(\beta IS+u(t)\right)-\lambda\left( I_{\mathrm{data}}-I\right)\\
\dot{I}&=\beta IS-\gamma I+u(t)\left(1-\frac{\gamma}{S\beta}\right),\\
\label{scalar_sir}
\end{aligned}
\end{gathered}
\end{equation}
where we omit for simplicity the constraint on the distribution of the parameters. The function $u(t)$ is a constraint ensuring that
the invariant of the model is satisfied,
\begin{equation}
    I+S-S_{0}-I_{0}-\frac{\gamma}{\beta}\mathrm{log}(\frac{S}{S_{0}})=0.
\end{equation}

The Lagrangian functional for the mcSIR model is more elaborate and can be derived in a similar manner. Similar to the Lagrangian of the SIR model (Eq. (\ref{eq:LSIR})), we derive the reduced Lagrangian of the mcSIR model, $L_{\mathrm{mcSIR}}(\dot{\mathbf{S}},\mathbf{S},\dot{\mathbf{I}},\mathbf{I})$. 
The functional is as follows: 
\begin{multline}
L_{\mathrm{mcSIR}}=L_{\mathrm{mcSIR}0}+\underset{i}{\sum}\int^t_0\frac{\dot{S}_{i}I_{i}-S_{i}\dot{I}_{i}}{2}dt\\
+\underset{i}{\sum}\int^t_0\left(\underset{j}{\sum}\beta_{ij}I_{j}S_{i}+u_{i}(t)\right)\left(
   I_{i}+S_{i}-S_{i0}-I_{i0}
    -\frac{
    \gamma_{i} I_i
    }{\underset{j}{\sum}\beta_{ij}I_{j}}\mathrm{log}(\frac{S_{i}}{S_{i0}})
\right)dt
\\+\underset{i}{\sum}\frac{\lambda_{i}}{2}\int^t_0\left( I_{\mathrm{data},i}-I_{i}\right)^2dt.
    \label{eq:LmcSIR}
\end{multline}
Where $u_i(t)$ is Lagrange multiplier function, enforcing a constraint on the invariant of the equation:
\begin{equation}
    I_{i}+S_{i}-S_{i0}-I_{i0}
    -\frac{
    \gamma_{i} I_i
    }{\underset{j}{\sum}\beta_{ij}I_{j}}\mathrm{log}(\frac{S_{i}}{S_{i0}})=0.
\end{equation}
The weights $\lambda_i$ are the Lagrange multipliers of each country $i$. The values are estimated during the optimization. The resulted weights are indication of the goodness of the fit to the data for each country. The Euler-Lagrange equations for the infected population in country $i$, $I_i$, are as follows: 
\begin{equation}
    \frac{d}{dt}\left(\frac{\partial L}{\partial\dot{I_i}}\right)=\frac{\partial L}{\partial I_i}
\end{equation}
Substituting the Lagrangian in Eq. (\ref{eq:LmcSIR}), one arrives to the following equations for the susceptible in country $i$ with an additional constraint on the similarity to the data:
\begin{multline}
\dot{S}_{i}=-\left(\underset{j}{\sum}\beta_{ij}I_{j}S_{i}+u_{i}(t)\right)
\\-\beta_{ii}S_{i}\left(
   I_{i}+S_{i}-S_{i0}-I_{i0}
    -\frac{
    \gamma_{i} I_i
    }{\underset{j}{\sum}\beta_{ij}I_{j}}\mathrm{log}(\frac{S_{i}}{S_{i0}})
\right)\\
=-\underset{j}{\sum}\beta_{ij}I_{j}S_{i}-u_{i}(t)-\lambda_i\left( I_{\mathrm{data},i}-I_{i}\right)
\end{multline}
Given this equation, we can find an expression for the functions, $u_i(t)=-\dot{S}_{i}-\underset{j}{\sum}\beta_{ij}I_{j}S_{i}$, at the stationary point. Similar to the infected population. The Euler-Lagrange equations for the susceptible population in country $i$, $S_i$, are as follows:
\begin{equation}
    \frac{d}{dt}\left(\frac{\partial L}{\partial\dot{S_i}}\right)=\frac{\partial L}{\partial S_i}
\end{equation}
Substituting the Lagrangian in Eq. (\ref{eq:LmcSIR}), one gets the following equation for the infected people in country $i$: 
\begin{multline}
\dot{I}_{i}=\underset{j}{\sum}\beta_{ij}I_{j}\left(I_{i}+S_{i}-S_{i0}-I_{i0}-\frac{\gamma_{i}}{\beta_{ii}}\mathrm{log}(\frac{S_{i}}{S_{i0}})\right)+\underset{j}{\sum}\beta_{ij}I_{j}S_{i}\left(1-\frac{\gamma_{i}}{\beta_{ii}S_{i}}\right)\\+\frac{\gamma_{i}}{\beta_{ii}}\underset{j\neq i}{\sum}\beta_{ij}I_{j}+u_{i}(t)\left(1-\frac{\gamma_{i}}{\beta_{ii}S_{i}}\right)\\=\underset{j}{\sum}\beta_{ij}I_{j}S_{i}-\gamma_{i}I_{i}-\left(\frac{\gamma_{i}}{\beta_{ii}}\underset{j\neq i}{\sum}\beta_{ij}I_{j}\right)+\frac{\gamma_{i}}{\beta_{ii}}\underset{j\neq i}{\sum}\beta_{ij}I_{j}+u_{i}(t)\left(1-\frac{\gamma_{i}}{\beta_{ii}S_{i}}\right)\\=\underset{j}{\sum}\beta_{ij}I_{j}S_{i}-\gamma_{i}I_{i}+u_{i}(t)\left(1-\frac{\gamma_{i}}{\beta_{ii}S_{i}}\right)
\end{multline}

Note that the advantage of using the Lagrangian framework in the mcSIR model is that, a-priory, we do not filter the time series of countries with noisy data. We run the optimization algorithm using all available data. The algorithm finds a solution based on the existing data, giving lower weight to noisy countries. In addition, solving the mcSIR model using the Lagrangian framework makes it possible to add a constraint on the distribution of the off-diagonal infection rates, such as $L_1$ norm on each row of the infection rate matrix. Such a sparsity constraint can account for spatial connectivity by limiting each country to interact with only a small number of countries.   

\section{Results\label{Sec:Results}}

Though there are $24*7*6=1008$ possible time series, in practice only 611 out of them contained sufficient data. The threshold set for the data to be considered sufficient is that the maximum of infection rate is greater than 1\% during one week of data and that the total percent of infection (per week) over the entire season was greater than 20\%.

The algorithm reached an average Pearson correlation between the estimated number of infected people according to the model and the WHO data of $r=0.59$ for SIR and  $r=0.57$ for mcSIR.  

\begin{figure}[bt]
\centering
\includegraphics[trim={4cm 11cm 4cm 10cm},clip,scale=0.7]{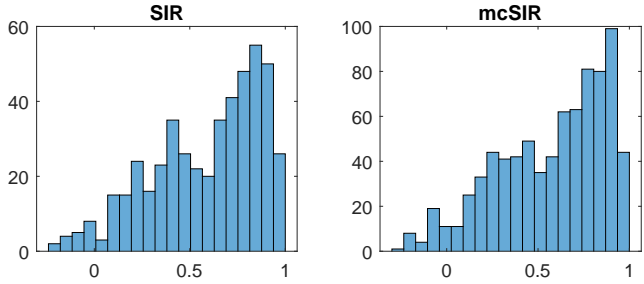}
\caption{\label{fig:HistogarmStateFit} An histogram of the Pearson correlation between the predicted number of infected people according to the models in $369$ time series over different strains and seasons and the number of infections according to WHO.}
\end{figure}

The mcSIR model fit is performed for all time series including ones below the threshold of significance. The algorithm learns smaller weights (Lagrange multipliers) for the time series that are very noisy and for ones with significant missing data. Figure \ref{fig:HistogarmStateFit} shows the distribution of the correlation between the WHO data and the model predictions for 6 strains and 7 seasons, for the SIR and the mcSIR models in each country. 

To test the assumption that the mcSIR Lagrangian assigns a lower weight to noisier data, we calculate the correlation between the quality of the fit achieved, i.e., one over the mean square error between the data and the estimated fit, and the weights estimated from the mcSIR Lagrangian framework. The Spearman correlation between the weights and the quality of the fit for the SIR model is, on average, $\rho=0.47$ and $\rho=0.52$ for the mcSIR model. 

\subsection{\label{sec:similarity}Infection and Recovery Rates in Different Strains and Seasons}

To ascertain whether there are similarities between the inter-country estimated infection and recovery rates for each strain at each season, we calculate the correlations between parameter pairs of different seasons and strains. Each item in a pair is a vector of the inter-country parameters. Parameter pairs are considered only if they have a weight (Lagrange multiplier) greater than $1e-3$. The values of the weights are within the range $0$ to $0.5$. The average fraction of countries that had a weight greater than $1e-3$ is $0.84$.

\begin{figure}[bt]
\begin{center}
\includegraphics[trim={2cm 7cm 0cm 7cm},clip,scale=0.6]{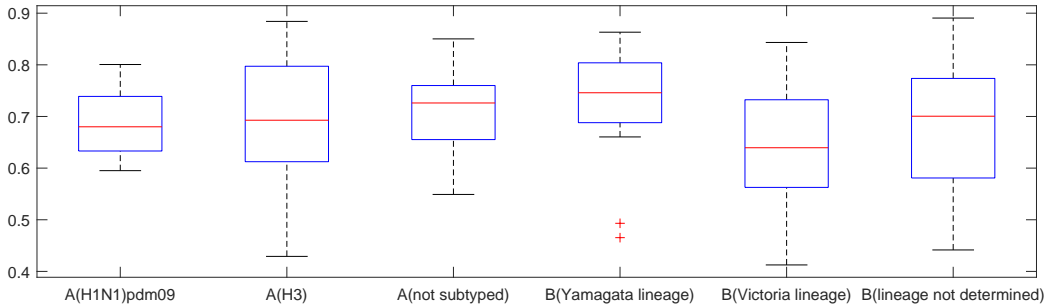}
\end{center}
\caption{\label{fig:Strians Pairs}
A box plot of the Spearman correlation between the infection rates and recovery rates for each pair of seasons of the same strain. On each box, the central mark indicates the median, and the bottom and top edges of the box indicate the 25th and 75th percentiles, respectively. The whiskers extend to the most extreme data points not considered outliers, and the outliers are plotted individually using the '+' symbol.
}
\end{figure}

\begin{figure}[bt]
\begin{center}
\includegraphics[trim={2cm 7cm 0cm 7cm},clip,scale=0.6]{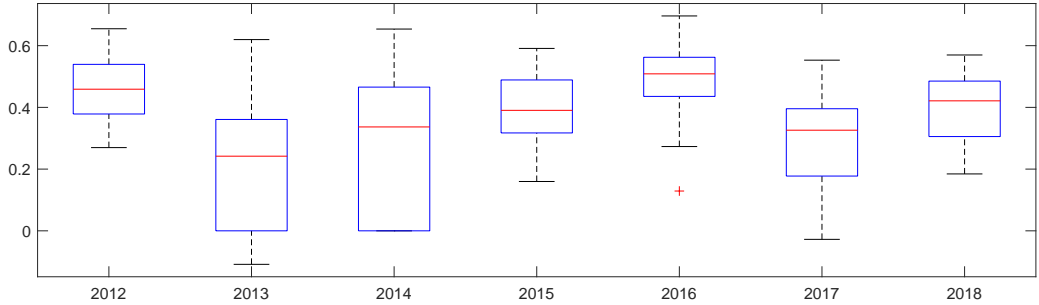}
\end{center}
\caption{\label{fig:Years Pairs}A box plot of the Spearman correlation between the infection rates and recovery rates for each pair of strains in the same season. On each box, the central mark indicates the median, and the bottom and top edges of the box indicate the 25th and 75th percentiles, respectively. The whiskers extend to the most extreme data points not considered outliers, and the outliers are plotted individually using the '+' symbol.
}
\end{figure}

Figure \ref{fig:Strians Pairs} shows a box plot of the Spearman correlation between pairs of parameters extracted in different seasons for the same strain and Figure \ref{fig:Years Pairs} shows the same among pairs of parameters extracted for different strains in the same season. As the Figures show, disease parameters among different strains and seasons are correlated. This fact is in agreement with surveillance of the spreading of influenza which indicates a similarity among seasons and strains\cite{WHOData,ginsberg2009detecting,ECDC}. We note that there is a higer correlation between diseses parameters among diffrent seasons than among diffrent strains. In the next section we show features of these correlations.  

\subsubsection{Parameter Similarity Among Influenza Strains}

Figures \ref{fig:Strians Pairs} and \ref{fig:Years Pairs} show that the estimated infection rates and recovery rates are correlated over seasons and strains. In this subsection, we further investigate the similarity between strains. 

We use the inferred on-diagonal disease parameters to cluster influenza strains using hierarchical tree of the average Euclidean distance between clusters of data points.

\begin{figure}[bt]
\begin{center}
\includegraphics[trim={2cm 1cm 1cm 1cm},clip,scale=0.55]{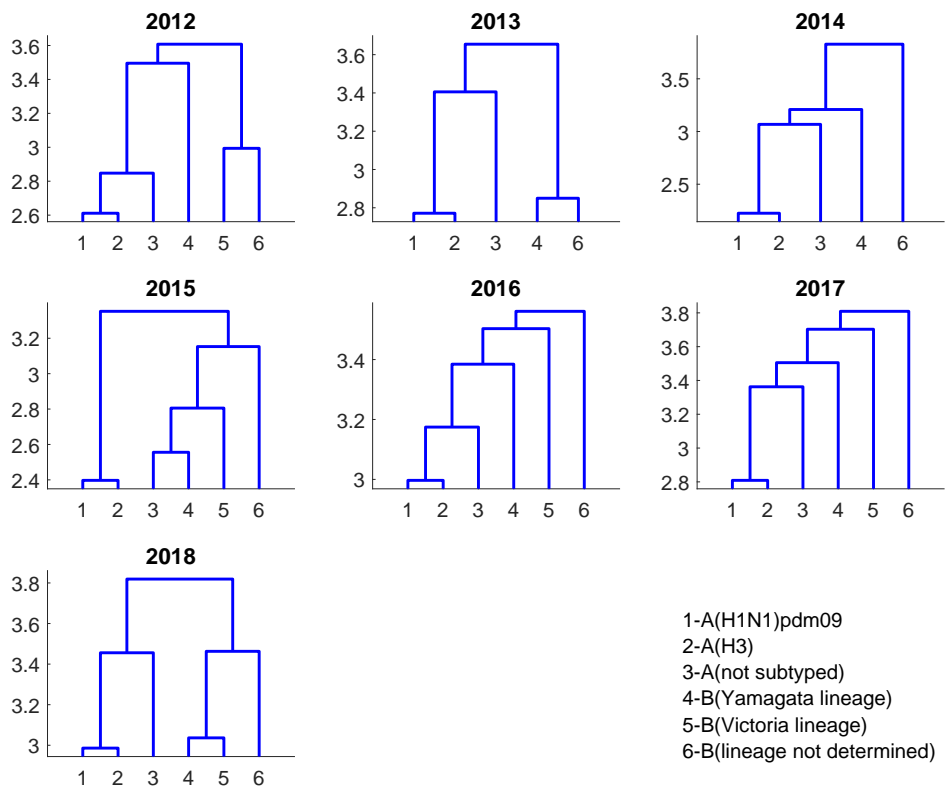}
\end{center}
\caption{\label{fig:Dendogram plots} Hierarchical plots of clustering 6 influenza strains according to the on-diagonal coefficients in different years. }
\end{figure}

Figure \ref{fig:Dendogram plots} presents dendogram plots of this clustering over seven seasons. As the Figure demonstrates, the distances between the disease parameters estimated using the mcSIR model matches the known virological classification of the influenza virus into types. Specifically, the clustering in almost all seasons shows a separation between influenza type A and type B. We also note the similarity among A(H1N1)pdm09 and A(H3) which is present in all seasons. 

\subsubsection{\label{Sec:HumanMobility}Human Mobility and Disease Parameters}

Each element of the off-diagonal infection rate matrix in the mcSIR model (Eq.(\ref{matrix_sir})) describes the rate at which infected people in one country infect people in another country. A previous study \cite{seroussi2019multi} showed that the estimated mobile susceptible people partially explains human mobility in the US. The estimated number of mobile susceptible people are defined by the off-diagonal infection rates multiplied by the average number of infected people during an entire season. In that study, human mobility in the US explained 30\% of the estimated number of mobile susceptible individuals \cite{seroussi2019multi}. 

There, to estimate the correlation between the state-level mobility patterns estimated from Twitter data and the values of $\bm{\beta}$, the matrix $\bm{\beta}$ is normalized by the average number of infected people estimated by the mcSIR model in each country during a particular season, i.e., the infection rate matrix $\bm{\beta}$ is multiplied by a diagonal matrix of the average number of people transmitting the disease in a season in each country. This quantity is defined as
the average mobility of the susceptible individual among different countries.

The same analysis performed on the influenza data in Europe reveals no correlation between the estimated normalized infection rates and human mobility as estimated from Twitter and as given in train and air transportation statistics. Table \ref{Tab:HumanMobility} presents the relevant Spearman correlations.

Note that the Spearman correlation between the Twitter matrix and air mobility is on average (over 6 years) $0.62$. On the other hand, the Spearman correlation between the Twitter matrix and train mobility is on average (over 6 years) $0.1$. 

\begin{table}[bt]
\caption{\label{Tab:HumanMobility}Average Spearman correlation of the estimated mobility of susceptible people, as estimated by the mcSIR model fit to the data for each strain over $7$ seasons. The p-value is above $0.05$ in all strains, apart from three cases marked by *.
}
\begin{tabular}{lccc}
& {\bf Air mobility} &{\bf Train mobility}&{\bf Twitter data}\\%\thickhline
\hline
A (H1N1)pdm09 & 0.05&-0.00& 0.07 \\
A (H3) &0.03&-0.01& 0.09\\
A (not subtyped) &0.38* & -0.13&0.34*\\
B (Yamagata lineage)&0.23*&0.04&0.1\\
B (Victoria lineage)&0.09&0.04&0.06 \\
B (lineage not determined) &0.00&0.00& 0.1\\
\hline  
\end{tabular}
\end{table}
\subsubsection{\label{Sec:HumanMobility}Data Completion using mcSIR}

One of the advantages of the mcSIR model over the SIR model is the former's ability to predict the number of infected people in one country based on data from adjacent countries. In this section we demonstrate this capability by predicting the infection rate in one country based on data from the surroundings countries in the current season together with disease parameters from the country from previews season. 

First, we simulate this ability (see Supporting Information \ref{SISec:data completion} for more details). Second, we show that this idea can be implemented with real data given some prior information on the infection and recovery rates in the country containing missing data from previews seasons. Note that this is possible even though human mobility seems not to play a significant role in prediction of the infection rate in the missing country. 

As an example, we simulated missing data for the United Kingdom of the influenza A (not subtyped) strain in the $2015$ season. We applied the mcSIR model to data from the $24$ countries, removing completely data from the United Kingdom. We initialized the number of infected people in this country to be zero. The on-diagonal infection and recovery rates were initialized to the ones approximated from the previous season. 

\begin{figure}[bt]
\begin{center}
\includegraphics[trim={5cm 10cm 5cm 10cm},clip,scale=0.9]{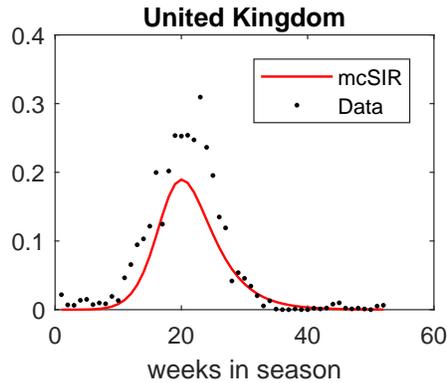}
\end{center}
\caption{\label{fig:UKdata} Estimated infection rates in the case of missing data. In red is the mcSIR-estimated infection rate in the United Kingdom, given infection rates in the other $23$ countries. In dotted black is the reported infection rate in the United Kingdom.}
\end{figure}

The resulting infection rate time series estimated based on the data from other countries (without the United Kingdom) are presented in Figure \ref{fig:UKdata}. The Pearson correlation between the WHO data (which was not available to the model) and the mcSIR-predicted rate was $0.96$ (p-value: $=5.5 \cdot 10^{-28}$). 

\begin{figure}[bt]
\begin{center}
\includegraphics[trim={4cm 10cm 4cm 10cm},clip,scale=0.7]{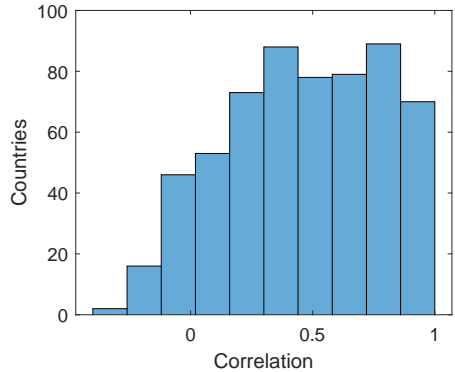}
\end{center}
\caption{\label{fig:MissingCountryCorr2015} Histogram of the correlations between the actual infection rate in the removed country and the mcSIR-predicted infection rate. One country is removed at each time. The prediction is performed for all strains in the seasons 2013 to 2017.}
\end{figure}

We then repeated this simulation of missing data by rerunning the model, each time removing one country, for all strains and for all seasons from 2013 to 2017. Figure \ref{fig:MissingCountryCorr2015} shows an histogram of the correlation between the predicted infection rate (over time) in the missing country and the actual infection rate. The average correlation is $0.46$ (s.d. $0.32$). 

The Figure shows that in most cases the model obtains a relatively high-quality prediction for the infection rate in the missing country based on neighbouring countries. Note that the distribution has a relatively heavy tail. One reason for this is that in some strains there is not enough data about other countries to provide a good predictions of the missing country. Additionally, we initiate the algorithm based on the average of the parameters over all previews seasons. Given that there is some variability between seasons (see Section \ref{sec:similarity}), additional work is required to to ascertain the number of previous seasons that should be used to estimate off-diagonal parameters. 

\section{Discussion}

Modeling the combined spreading mechanism of multiple circulating pathogens in multiple geographic regions and among multiple populations is a challenging problem because of the dependencies between different factors affecting these regions, pathogens, and populations. In addition, the available epidemiological data is usually noisy and incomplete. Here we analyze data on seasonal influenza of multiple strains and in different countries in Europe. In order to exploit all available information and account for low-quality data we apply a new Lagrangian optimization framework. This framework enables us to estimate the infection and recovery rates using the mcSIR model while weighting differently the data from different countries. Our novel framework is general and can be applied easily to other compartment models.

Our analysis shows the importance of modeling heterogeneity in Europe, as our estimated infection and recovery rates differ for each country, as opposed, for example, to the US \cite{seroussi2019multi}, where such differences are smaller. The parameters extracted are not independent: We show significant correlation between the inter-country infection and recovery rates in different seasons ($~0.6-0.75$). In addition, we show that the distances among the disease parameters in different strains has an hierarchical structure which is consistently preserved over many season. This hierarchical structure matches genomic similarity among influenza subtypes, lending additional support to the value of these estimated parameters. In addition, as we show in the simulation, the Lagrangian framework in the mcSIR model allows us to estimate the infection rate in a country even when there is no available data on this country, simply from data of adjacent countries. 

Surprisingly, the off-diagonal infected rates extracted in each season and strain are very small and uncorrelated with human mobility. This is in opposition to our finding about the significant role of human mobility in the infection mechanism of the Respiratory Syncytial Virus and West Nile Virus in the US \cite{seroussi2019multi}. Previous work has suggested that mobility across countries in Europe has a negligible role in disease transmission \cite{merler2009role}. Here we find, for the first time, a direct indication that inter-country human mobility does not play significant role in the spreading of influenza over many strains and seasons of the influenza virus in Europe. 

Potential limitations of this study are in the model we use, which makes significant assumptions. These include the assumption that there is no mixing interaction between strains which is known to affect the dynamics and the equilibrium state \cite{garmer2015multistrain}. Moreover, our analysis do not addresses different social and economic groups, such as age group, which were shown to play significant role \cite{apolloni2014metapopulation}. We did not account for differential vaccination rates between countries and seasons. In addition, we are limited by the resolution of the data i.e., the reported number of infected people in each country.  Finer scaled data on the reported number of infected people within cites and counties may reveal more information about the spatial spreading mechanism.  Future work will focus on modeling such sub-populations within a geographic area, as well as the effects of interventions such as vaccination. 
\section{Materials and methods}
\subsection{Data Sources}
\subsubsection{World Health Organization (WHO) Surveillance Data}
The WHO produces surveillance data for a number of emerging diseases such as Zika, Ebola, and influenza \cite{WHOData}. We use the weekly reported number of influenza A and B viruses classified by sub-type in Europe over the years 2012 to 2018 in the following $24$ countries: France, Georgia, Germany,  Spain, Italy, Austria, Denmark, Iceland, Hungary, Sweden, Turkey, Netherlands, Norway, Belgium, Portugal, Poland, Greece, Finland, Romania, Ireland, Czechia, Slovakia, Bulgaria, United Kingdom. 

We define the influenza season to be one year starting at the beginning of week 41 of each year. 

The sub-type taken are A (H1N1)pdm09, A (H3), A (not subtyped), B (Yamagata lineage), B (Victoria lineage), B (lineage not determined). 

\subsubsection{Human Mobility Data}

We extracted data on air and train mobility among different countries in Europe between the years 2012 to 2017 from the Eurostat website \cite{EurostatData}. 

Additionally, we collected all messages from Twitter having a GPS location in Europe from October 2015 through to March 2016. For each message, we extracted an
anonymous user identifier, the time of the message, and the location from where
it was made. These data comprised of approximately 50 million messages and 1.2 million users. The exact GPS location of each message was mapped as its encompassing European country.
We then create a matrix of mobility, where the $(i, j)$-th entry of the matrix
comprised of the number of people whose location in one tweet was country $i$ and in their following tweet was country $j$. This matrix was normalized by dividing the number of people moving from country $i$ to country $j$ by the total number of people who moved from country $i$ to any other country. 

The Twitter data provide an estimate for the total movement between countries in Europe. The advantage of Twitter data over the Eurostat data is that the former provide an estimate for total human mobility, including, for example, roads, which are known to be the main source of transportation in Europe \cite{EurostatData}. 
\newpage
\appendix
\section*{Appendix}
\section{Fitting the Model to Observed Data}

The optimization procedure is performed in two steps: First, we separately estimate the infection and recovery rates in each country, season and strain. This is done by using the Lagrangian framework of the SIR model as shown in Eq. (\ref{eq:LSIR}). We minimize the Lagrangian using gradient descent. The estimated parameters are used as an initialization of the on-diagonal elements of the infection and the recovery rates in the mcSIR model over multiple countries for each strain and season. We then find the optimal solution of the mcSIR model using the mcSIR Lagrangian in Eq. (\ref{eq:LmcSIR}). The off-diagonal terms are initialized to be the average value of the infection rates within countries divided by the total number of countries. Thus, to fit the model, $(24^2+24)*6*7$ parameters are estimated. 

The Lagrangian optimization process is performed using an alternating minimization procedure. We start with an initial guess for the fraction of infected people based on the data. We estimate the corresponding susceptible population using the Euler-Lagrange equations. The susceptible and the infected people time series found are used to minimize the Lagrangian and estimated the optimal parameters of the model. The minimum is found using the function \textit{fminsearch} in Matlab. Using these parameters, we find a new estimate for the infected population by using again the Euler-Lagrange equations. We iterate the procedure until convergence. The advantage of using the Lagrangian framework for the mcSIR model is that it provides a way to assign smaller weights to countries with noisy data in a controlled manner while taking into account all the data available. 
All analyses were performed using Matlab R2017b \cite{MatlabSMLTB}.

\section{Fitting data using a Particle Swarm Algorithm \label{SISec: PS}}
In order to validate numerically that our results are independent on the initial conditions, we also estimated the infection and recovery rates using an additional optimization algorithm which does not depend on the initial conditions of the single SIR model. This analysis also shows the strength of the Lagrangian framework in filtering the noisy data. 

The algorithm we use is the particle swarm algorithm \cite{eberhart1995particle}. Particle swarm optimization is a stochastic population-based optimization method proposed by Kennedy and Eberhart \cite{eberhart1995particle}. 

We find that the results obtained using this algorithm are similar to those of the Lagrangian framework: The algorithm reached an average Pearson correlation $0.57$ for SIR and $0.68$ for mcSIR. There are $24*7*6=1008$ time series data points in practice only $287$ out of them contained significant data, defined as a maximum of infection greater than $0.01$ and a total infection over the season greater than $0.2$. 

\begin{figure}[hbt!]
\centering
\includegraphics[trim={4cm 11cm 4cm 10cm},clip,scale=0.7]{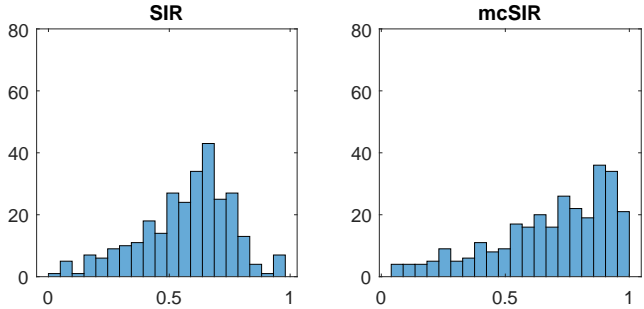}
\caption{\label{S1fig:HistogarmStateFitPS} An histogram of the Pearson correlations achieved between the predicted infection rate in each country and WHO data, using the SIR and the mcSIR models using swarm optimization.}
\end{figure}

\begin{figure}[hbt!]
\begin{center}
\includegraphics[trim={2cm 7cm 0cm 7cm},clip,scale=0.6]{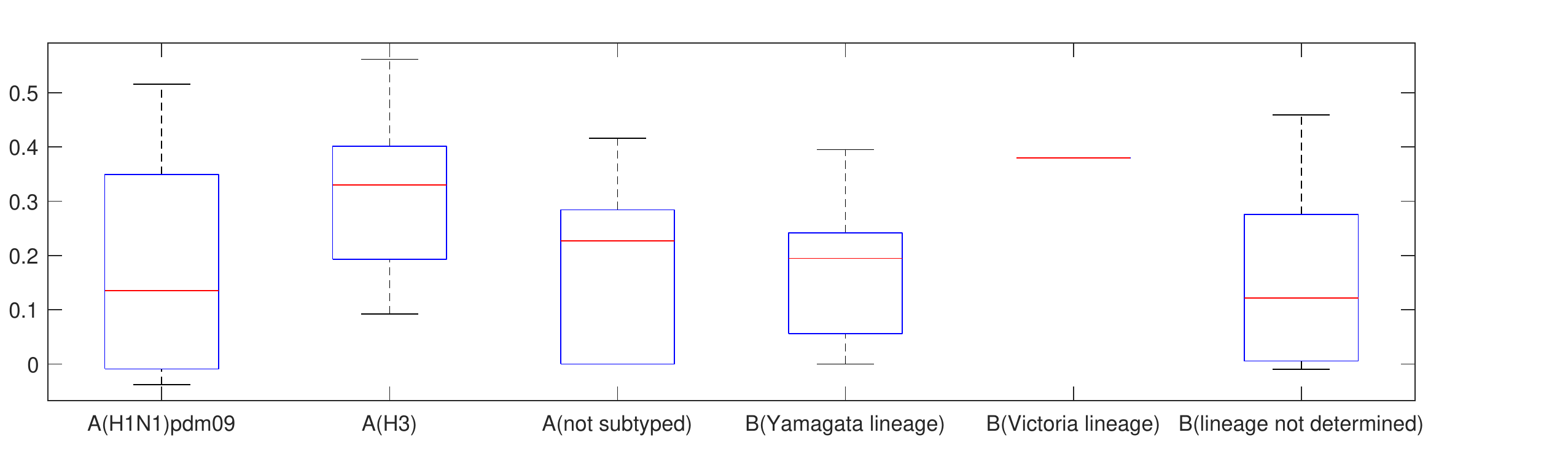}
\end{center}
\caption{\label{S2fig:Strians PairsPS} A box plot of the Spearman correlation between the infection rates and recovery rates for each pair of seasons of the same strain, as found using swarm optimization. On each box, the central mark indicates the median, and the bottom and top edges of the box indicate the 25th and 75th percentiles, respectively. The whiskers extend to the most extreme data points not considered outliers, and the outliers are plotted individually using the '+' symbol.}
\end{figure}

\begin{figure}[hbt!]
\begin{center}
\includegraphics[trim={2cm 7cm 0cm 7cm},clip,scale=0.6]{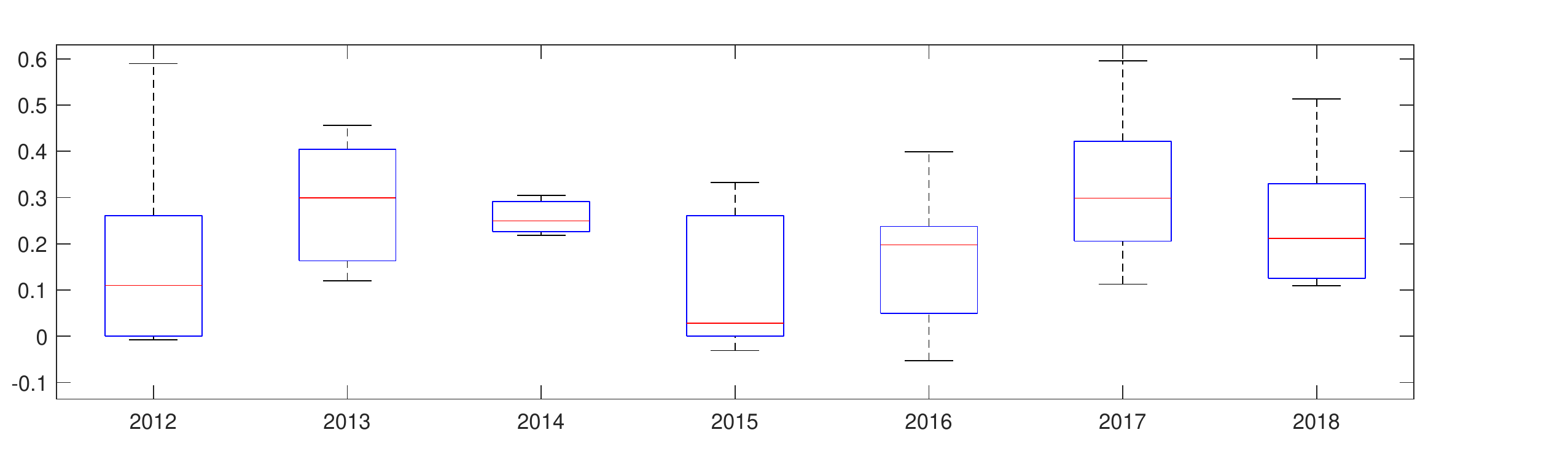}
\end{center}
\caption{\label{S3fig:Years PairsPS} A box plot of the Spearman correlation between the infection rates and recovery rates for each pair of strains in the same season, as found using swarm optimization. On each box, the central mark indicates the median, and the bottom and top edges of the box indicate the 25th and 75th percentiles, respectively. The whiskers extend to the most extreme data points not considered outliers, and the outliers are plotted individually using the '+' symbol.}
\end{figure}

\begin{figure}[hbt!]
\begin{center}
\includegraphics[trim={0cm 1cm 0cm 1cm},clip,scale=0.5]{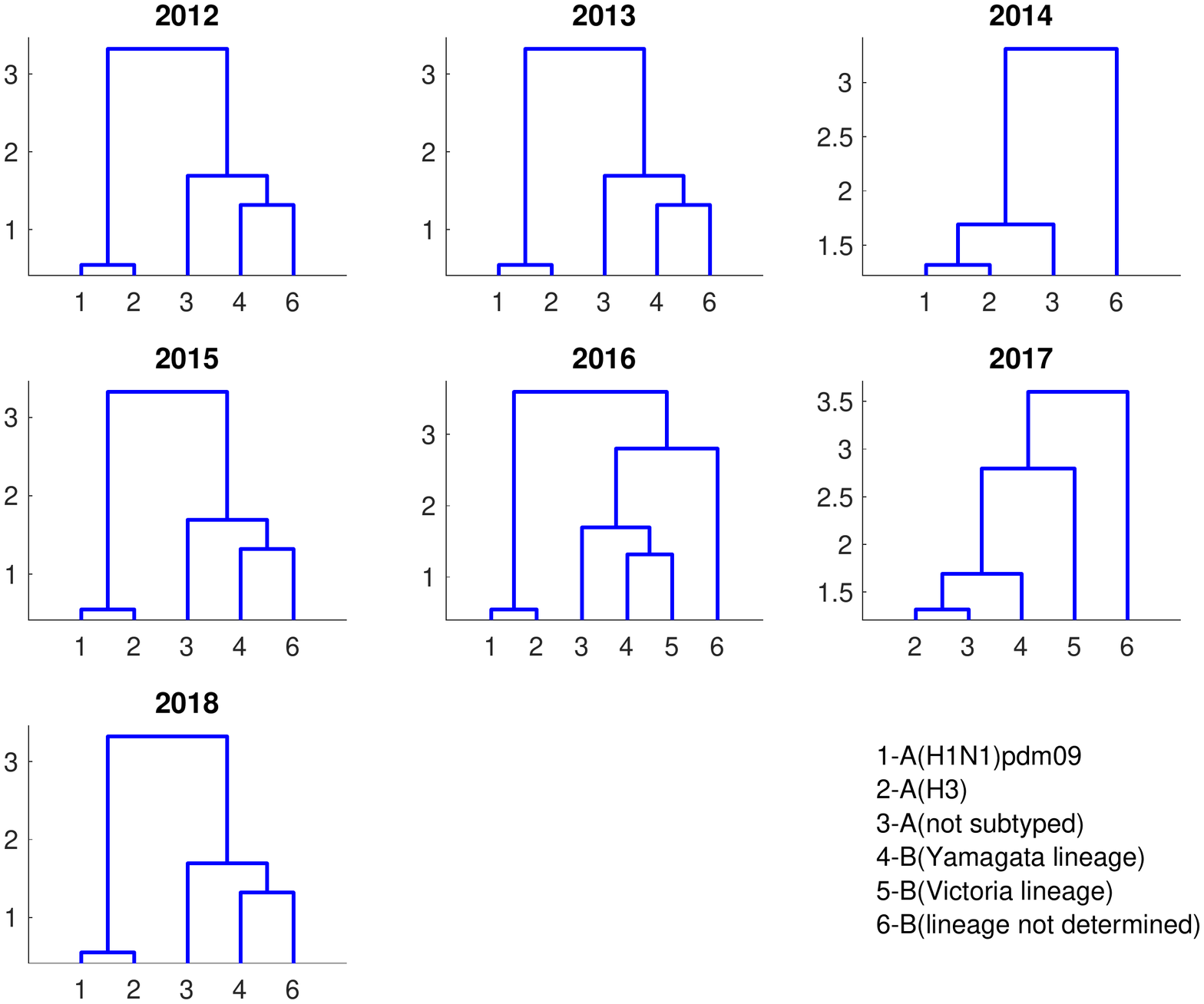}
\end{center}
\caption{\label{S4fig:Dendogram plotsPS} Linkage plots of the average Euclidean distance of the on diagonal coefficients in different years over 6 strains. The linkage tress shows separation in the estimated coefficients between strain A and strain B.}
\end{figure}

Figures \ref{S1fig:HistogarmStateFitPS}, \ref{S2fig:Strians PairsPS}, \ref{S3fig:Years PairsPS}, and \ref{S4fig:Dendogram plotsPS} present the results. These results are with agreement with our Lagrangian framework showing correlation between the parameters in different seasons and strains. See Sec. \ref{Sec:Results} in the main text. Note that, using the Lagrangian framework we reach better results, mainly due to the weighting mechanism provided by the Lagrange multipliers. This mechanism allows us to filter the noise in the data while learning the model parameters. 
  

\section{mcSIR model completion of data based on neighboring countries \label{SISec:data completion}}

In this section, the mcSIR model is used to complete missing data in countries where the data is missing, based on information on the number of infected people in neighbouring countries. We provide two results: First, using simulated data we show that mcSIR can be used to complete missing data points. Second, using real data we demonstrate the effectiveness of the mcSIR model in predicting the infection rates in countries with missing data. The results on the real data are presented in Sec. \ref{Sec:HumanMobility} in the main text. 

We used the mcSIR model to simulate synthetic infection rates of $6$ countries. $\gamma_i$ and $\beta_i$ were assigned uniformly random values for each country $i$ in the range of $0-2$. The interaction between country $i$ and $j$ is taken to be $\beta_{ij}=\beta_i J_{\mathrm{int}}$, and $\beta_{ii}=\beta_i (1-J_{\mathrm{int}})$, where $J_{\mathrm{int}}$ is a number between zero and one. 

To simulate data quality issues, we removed random points from the data and added Gaussian noise. We also removed completely the data from one country. 

\begin{figure}[hbt!]
\begin{center}
\includegraphics[trim={0cm 7cm 0cm 7cm},clip,scale=0.5]{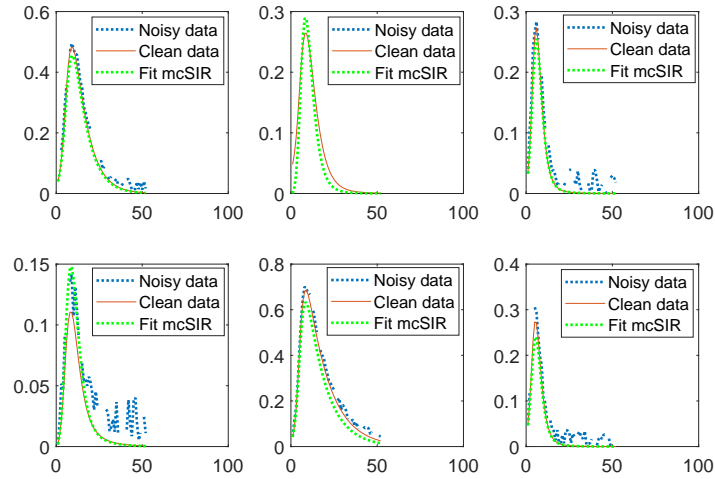}
\end{center}
\caption{\label{S6fig:simMissingData} Results of the model fitting to mcSIR simulation over $6$ countries. The data from the first state is missing. In red is the clean simulated infected rate. The green curve is the estimated solution using the mcSIR and in dotted blue is the noisy synthetic data.}
\end{figure}

The mcSIR framework with the Lagrangian solver was used to estimate the number of infected people in all countries. 
Figure \ref{S6fig:simMissingData} shows the results of the simulation for $J_{\mathrm{int}}=0.1$, here we remove 10 random points sampled without replacement and added noise with a variance of $0.04$. The data from the first state was removed completely.  

\begin{figure}[hbt!]
\begin{center}
	\includegraphics[width=\textwidth,trim={2cm 7.5cm 2cm 8cm},clip,scale=1]{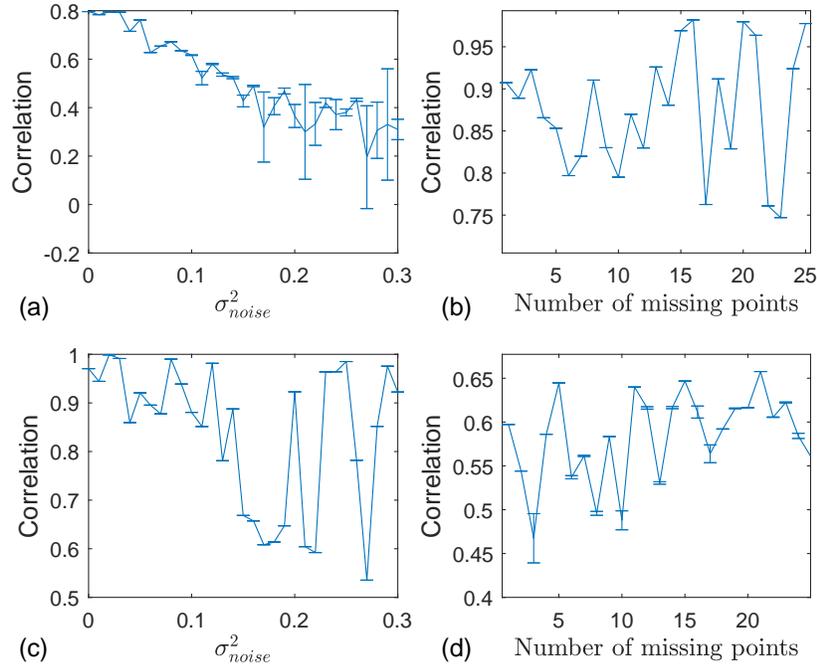}
\caption{\label{S5fig:simMissDataCorrNoiseTot} (a-b) The Pearson correlation between the noisy simulated data of $5$ countries and the estimated infection rate time series using the mcSIR for $6$ countries as a function of the variance of the added noise. In (a) the noisy data is generated by removing at random 25\% of the data points at each iteration and adding noise with different variance $\sigma^2_{\mathrm{noise}}$. In (b) the noisy data is generated by removing a different number of random points at each iteration and adding noise with variance $\sigma^2_{\mathrm{noise}}=0.1$. (c-d) The Pearson correlation between the estimated infection rate time series in the missing country using mcSIR for 6 countries and the clean data as a function of the variance of the added noise. In (c) the noisy data is generated by removing at random quarter of the point at each iteration and adding noise with variance $\sigma^2_{\mathrm{noise}}$. In (d) the noisy data is generated by removing at random different number of the points at each iteration and adding noise with variance $\sigma^2_{\mathrm{noise}}=0.1$. The data from the $6$ country is removed completely.}
\end{center}
\end{figure}

The average Pearson correlation between the simulated noisy data and the fitted signal was $r=0.98$. The Pearson correlation between the simulated clean signal data in the first state and the fitted signal was $r=0.97$. We performed the same experiment multiple times while each time removing a different set of points and increasing the signal to noise ratio. We notice that the algorithm is less sensitive to the number of points removed. On the other hand, a large perturbation in the initialization point can affect the prediction. Figure \ref{S5fig:simMissDataCorrNoiseTot}a shows the correlation between the predicted signal in $5$ countries on which the mcSIR optimization on $6$ countries was done on
as a function of the variance of the added noise.  Figure \ref{S5fig:simMissDataCorrNoiseTot}b presents the correlation between the predicted signal in $5$ countries on which the mcSIR optimization on $6$ countries was done on
as a function of the number of missing data point given noise variance $\sigma^2_{\mathrm{noise}}$.  
Figure \ref{S5fig:simMissDataCorrNoiseTot}c presents the correlation between the clean signal in the missing country and the predicted signal of this country using the mcSIR on $6$ countries while performing the optimization procedure given $5$ countries as a function of the variance of the added noise.  Figure \ref{S5fig:simMissDataCorrNoiseTot}d presents the correlation as a function of the number of missing data points given noise variance $\sigma^2_{\mathrm{noise}}$.
These results presents the robustness of the algorithm to two noise types. In addition, it shows the ability of the mcSIR model to complete for missing data based on data from neighboring countries. 

\newpage
\bibliography{multi_SIR}
\end{document}